\documentclass[doublecol]{epl2} 
\newcommand {\apgt} {\ {\raise-.5ex\hbox{$\buildrel>\over\sim$}}\ }

\title{Observation of metastable hcp solid helium.}
\author{F. Souris \and J. Grucker\thanks{E-mail: \email{grucker@lkb.ens.fr}} \and J. Dupont-Roc \and Ph. Jacquier}
\shortauthor{F. Souris \etal}

\institute{                    
   Laboratoire Kastler Brossel, \'{E}cole Normale Sup\'{e}rieure, Universit\'{e} Pierre et Marie Curie, CNRS, 24 rue Lhomond, 75005 Paris, France}

\abstract{We have produced and observed metastable solid helium-4 below its melting pressure between 1.1~K and 1.4~K. This is achieved by an intense pressure wave carefully focused inside a crystal of known orientation. An accurate density map of the focal zone is provided by an optical interferometric technique. Depending on the sample, minimum density achieved at focus corresponds to pressures between 2 and 4~bar below the static melting pressure. Beyond, the crystal undergoes an unexpected  instability much earlier than the predicted spinodal limit. This opens a novel opportunity to study this quantum crystal in an  expanded  metastable state and its stability limits.}

\begin{document}

\maketitle

%\section{Section title}
Crystalline solids can be brought into  metastable state with respect to fusion if surface melting is avoided. Experimentally this is achieved by heating small metal or dielectric crystals embedded in stronger matrices beyond the bulk melting temperature\cite{daeges1986,mei2007}. Metals are the most frequently studied materials. Fast heating of bulk solids by subpicosecond laser pulses  has also  provided  evidence of such  metastable states\cite{williamson1984,siwick2003}. Numerous computer simulations have been used to unveil mechanisms  limiting the metastability domain\cite{mei2007,bai2005,jin2001,belonoshko2006,luo2003}.
  
Solid helium offers novel opportunities.  At low temperatures ($0<T<1.4$ K), hcp solid helium-4 only exists for pressures higher than $P_m(T) \apgt 25$ bar. The melting line is nearly horizontal in the $T$-$P$ plane, in contrast with other simple solids where it is nearly vertical. In this unique situation, pressure is the efficient control parameter to explore the metastability region below the melting line. In addition, solid helium offers the peculiarity that quantum zero point motion plays a more important role than thermal motion. For melting, it does not obey the empirical  Lindemann criterion\cite{lindemann1910}\footnote{For hcp solid helium $\sqrt{<u^2>}/a\sim0.26$ at melting instead of 0.1. See\cite{burns1997}}. Metastable expanded solid helium-4 has also been considered in the past in view of achieving the Andreev-Lifshitz-Thouless scenario for supersolidity\cite{andreev1969,thouless1969} by lowering the vacancy energy\cite{pollet2008}.  Although this scenario is now considered less relevant\cite{boninsegni2006, balibar2010},  the interstitial and vacancy energies in a stretched lattice   has been computed\cite{boninsegni2006}, and a limit of the metastable region has been proposed to lie 35 bar below the melting line\cite{maris2009}.     
None of these prospects ever came to an experimental realization\footnote{Attempts to obtain metastable solid helium at much higher pressures and temperatures using temperature as the control parameter are reported in\cite{jung1987}} and a presumably large  domain for the solid metastable phase remains unexplored.

%*[{ }] [{. For hcp solid helium $\sqrt{<u^2>}/a\sim0.26$ at melting instead of 0.1. See }] lindemann1910, *burns1997
%*[{Attempts to obtain metastable solid helium at much higher pressures and temperatures using temperature as the control parameter are reported in~}][] jung1987

In this letter, we provide full experimental evidence that metastable solid helium-4 can be obtained well below $P_m(T)$ by a focused pressure wave inside bulk solid. This technique was used to study the liquid phase of helium in a metastable state with respect to its vapor or to the solid phase\cite{nissen1989,balibar2002}. For the hcp lattice of solid helium-4, the sound velocity is anisotropic. A pressure wave front originating from a point source is not a  sphere, but a prolate surface along the $c-$axis\cite{crepeau1972}. Hence hemispherical transducers do not produce focused sound waves  in that case. Decompression of hcp solid helium with an hemispherical transducer has actually been tried with only a modest success\cite{sourisJLTP2010}. Here we use an appropriate shaping of the ultrasound emitter (piezo-electric transducer PZT) to compensate for the sound velocity anisotropy\footnote{A cylinder of bulk ceramic PZT401 from Morgan Electroceramics was  grinded on a CNC lathe prior to metallization and poling. The inner surface matches half the wave surface of the quasi-longitudinal sound mode. The outer surface is parallel to the inner one providing a constant thickness of $2$ mm. The equatorial part is then cut away over $1$ mm in order to make the focal region visible.}. The density modulation, below and above the coexistence density, is accurately measured using an interferometric imaging technique.
 \begin{figure}[ht]
 \includegraphics[width=8cm,clip=]{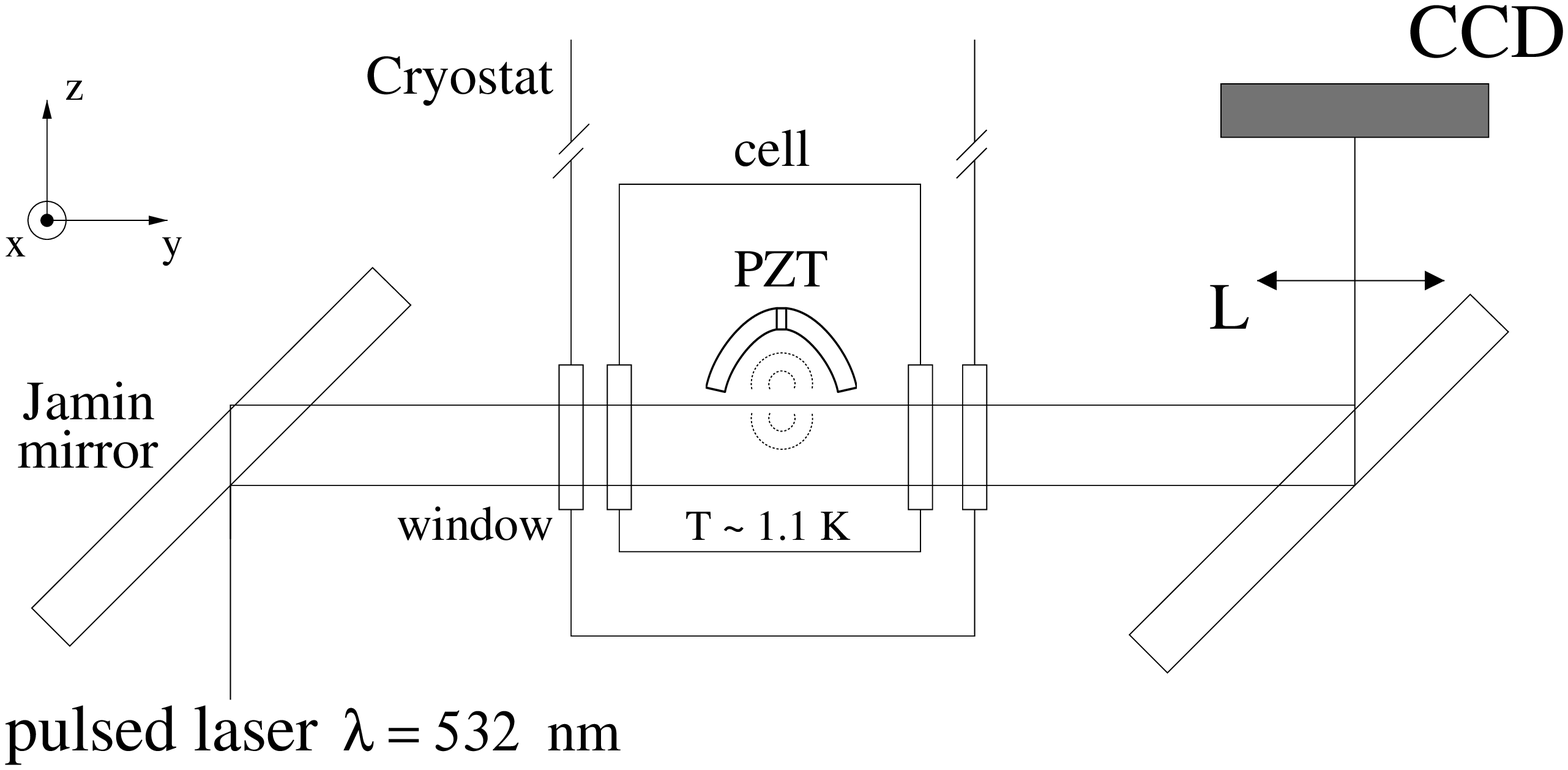}
\caption{A convergent pressure wave is produced in hcp solid helium at $1.1$ K by the excitation of a well shaped piezo-transducer (PZT). The density modulations near focus are monitored using a Jamin interferometer. The lens $L$ images the focal region onto the CCD camera}
\label{manip}
 \end{figure}

The experimental cell containing solid helium is cooled by a helium-4 cryostat with four optical ports. The temperature can be regulated from $1.1$ K to $2.1$ K. Internal pressure is measured by a capacitive gauge\cite{adams1993} calibrated along the liquid-solid coexistence line against an external pressure gauge\footnote{Keller PAA-35/30bar/80797}. The symmetry axis of the PZT is vertical and is taken as $z$-axis (Fig.~\ref{manip}). The orientation of solid helium should be such that its symmetry axis (perpendicular to the $c$-facet) coincide with  the $z$-axis. Helium density modulations are monitored via the refractive index using a Jamin interferometer (Fig.~\ref{manip}). One of its arms passes through the acoustic focal region and is taken as the $y$-axis, while the other one crosses the cell in an unperturbed zone. The focal region is imaged onto a CCD camera using an appropriate lens. The imaged surface is about $4\times2.4$ mm$^2$. The measurement technique has been described elsewhere\cite{souris2010} and may be summarized as follows: optical phase shifts $\varphi(x,z)$ induced by density modulations are extracted from several images taken with different phases of the reference beam. This process is then reproduced for different values of the delay $t$ between the sound pulse emission and the imaging laser pulse to construct a phase map movie. Because of the rotational invariance of the sound field around the  $z$-axis, the variation of the refractive index $\delta n$ depends only on the coordinate $z$ and on the distance $r$ to the $z$-axis. The optical phase map is then the integral along the $y$-axis of $\delta n(r,z)$, i.e.  its Abel transform. An inverse Abel transform\cite{deutsch1983} of $\varphi(x,z)$ provides the radial profile $\delta n(r,z)$. Density maps can be computed from $\delta \rho(r,z)=\rho_0\: \delta n(r,z)/(n_0-1)$ where $\rho_0$ and $n_0$ are the unperturbed density and refractive index of the solid. This method of determining sound fields has been quantitatively tested in water by comparison with an optical fiber hydrophone\cite{souris2010}. For an easier appreciation of the modulation achieved, conversion to pressure is made using:
\begin{equation}
\delta P(r,z)=\chi\: \frac{\delta n(r,z)}{n_0-1}%=\chi\: \frac{\delta \rho(r,z)}{\rho_0}
\label{eq_dpdn}
\end{equation}   
where $\chi=269$~bar\cite{grilly1973} is the bulk modulus\footnote{The use of formula (\ref{eq_dpdn}) is justified because solid helium compressibility is nearly isotropic ($c_{11}+c_{12}=c_{33}+c_{13}$). See\cite{wanner1970}} and $n_0=1.036$\cite{edwards1957}. This formula is  the linear approximation of the equation of state, valid up to $\delta P\simeq 5$~bar with an error less  than a few percent.  

%\cite{grilly1973,*[{The use of formula (\ref{eq_dpdn}) is justified because solid helium compressibility is nearly isotropic ($c_{11}+c_{12}=c_{33}+c_{13}$). See }] [{}] franck1970}
%*[{$n_0$ is computed using the Clausius-Mossotti relation and the polarizability determined by }] [{}] edwards1957

\begin{figure}[th]
 \includegraphics[width=8cm,clip=]{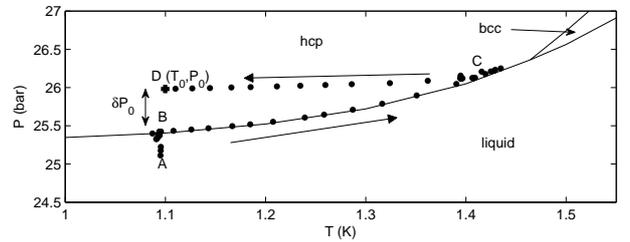}
 \caption{Experimental path in helium-4 phase diagram to obtain a crystal at $T_0$ and $P_0$. $A\rightarrow B$: liquid pressurization at constant $T$. $B$: nucleation of an oriented crystal seed. $B\rightarrow C$: warming of the crystal seed in equilibrium with the liquid up to $T_C\sim1.4$ K. $C$: growth of the crystal until the whole cell is filled. $C\rightarrow D(T_0,P_0)$: isochoric cooling. Coexistence curves are taken from Grilly\cite{grilly1973}. Solid dots are experimental data  provided by the inner pressure gauge. In this example, $\delta P_0=0.54$~bar.}% les donnes sont les fichiers pilchard: 2011Mar16_14h28.dat et 2011Mar17_08h42.dat
\label{chemin}
 \end{figure}

When producing solid under depression, nucleation of the liquid phase on the PZT's surface must be avoided. Let $\delta P_{PZT}(t)$ be the acoustic pressure at the PZT surface, $T_0$ the temperature of the cell and $P_m(T_0)$ the static melting pressure. We note $P_0$ the uniform pressure in the cell and $\delta P_0= P_0 - P_m(T_0)  $ its initial offset from the melting line (Fig.~\ref{chemin}). $P_0$ is chosen so that the condition $\delta P_{PZT}(t)+\delta P_0>0$ is always fulfilled, even at the troughs of the pressure wave. 

To obtain an oriented and homogeneously pressurized crystal at ($T_0,P_0$),  a crystal seed with its $c$-facet horizontal is first grown at a temperature below the roughening transition of the $c$-facet ($1.3$ K)\cite{balibar2005}. The horizontality of the facet is controlled optically in both $x$ and $y$ directions with 1$^{\circ}$ accuracy. Filling the cell with the crystal and pressurizing it from $P_m(T_0)$  to $P_0$ at constant temperature $T_0$ would be harmful for  crystal quality. Indeed, because helium is a very compressible solid, material would have to be brought into the cell including into the PZT cup. This would result in plastic deformation of the crystal,  possibly  breaking it into several differently oriented pieces. Pressure gradients could also be created. Instead we follow the procedure shown in  Fig.~\ref{chemin}. The cell with the seed in equilibrium with the liquid is warmed up to some temperature $T_C$ chosen so that $P_m(T_C)>P_0>P_m(T_0)$. This warming along the melting line is done slowly enough to keep a quasi-static equilibrium between the pressure in the filling line and the melting pressure at the cell temperature. The crystal is then grown at $T_C$ until the cell is filled and the filling capillary is blocked by the solid phase. Eventually, the whole cell is slowly cooled at constant volume down to $(T_0,P_0)$ in a little more than two hours. This slow isochoric cooling is expected to avoid crystal breaking, as well as temperature and pressure gradients.  

 \begin{figure}[ht]
 \includegraphics[clip, width=8cm]{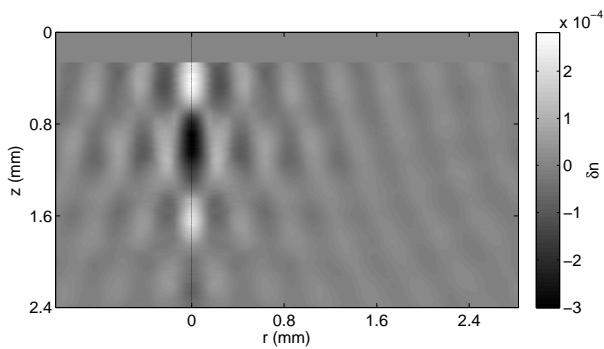}%trim= 25mm 90mm 15mm 85mm,
 \caption{Refractive index map $\delta n(r,z)$ of a convergent pressure wave in hcp helium single crystal. PZT voltage is $V=185$~V and propagation time from the PZT surface is $t=15.53$~$\mu$s. For an easier appreciation of the wave field around the focus, the map is supplemented on the left of the axis by a mere symmetry. } % les donnees sont sur backup: 2011-03-04_19h05m, Vem fit= 185V
\label{fig_indexmap}
 \end{figure}
\begin{figure}[ht]
 \includegraphics[clip, width=8cm]{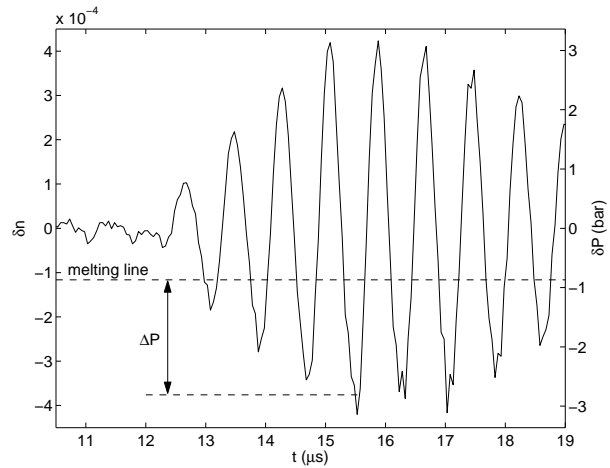}%trim= 15mm 70mm 5mm 65mm, 
 \caption{Time evolution of refractive index and pressure variations at focus for $V=185$ V. Maximum transient depression ($\delta P_{min}=-2.8$ bar $\pm0.2$ bar) is obtained at $t_{min}=15.53$ $\mu$s and is clearly below the static melting line indicated by the dotted line at $-\delta P_0=-0.87$ bar.}
%manip du 2011-03-04_19h05m, Vem=185V
\label{fig4}
 \end{figure}

The experiment then  runs as follows. A four-oscillation resonant RF pulse (frequency $1.25$ MHz, amplitude $V$ tunable from $10$ to about $250$ V) is applied to the PZT and generates a convergent pressure wave train in the crystal. Phase maps are recorded for delays $t$ ranging from $10\,\mu$s to  $20\,\mu$s with  $0.05\,\mu$s time steps. Radial refractive index maps $\delta n(r,z,t)$ are obtained from them. Such a map is shown in Fig.~\ref{fig_indexmap}. From those maps, one can find the acoustic focus $z_f$ as the location where $\delta n(0,z,t)$ has its maximum amplitude. In Fig.~\ref{fig4}, the refractive index variation at acoustic focus $\delta n(0,z_{f},t)$ is plotted versus time. The pressure scale of the right axis is computed using equation (\ref{eq_dpdn}). The pressure reaches its minimum at $t_{min}=15.53$ $\mu$s and its variation from the static pressure $P_0$ is $\delta P_{min}=-2.80$~bar $\pm0.12$~bar. The error bar is computed from the r.m.s. noise on $\delta n$ and corresponds to one standard deviation.
The initial offset pressure being $\delta P_0=0.87$~bar, one concludes that the solid is decompressed under its melting pressure by $\Delta P=\delta P_{min}+\delta P_{0}=-1.93\pm0.12$~bar. In another run with $\delta P_0=0.23$~bar, the value $\Delta P=-2.2\pm0.05$~bar was obtained.% c'est DPm des 4 mars et 31 mars

Does helium stay in the solid phase? If a liquid bubble had nucleated at the focus, it would have expanded for about a quarter of the sound period at a velocity close to the sound velocity. Thus it would have reached a diameter of the order of $100$ $\mu$m. Taking into account the refractive index difference between liquid and solid phases (about $0.0035$), the liquid bubble would have created a phase shift as large as 4 radians on a size larger than the $20$ $\mu$m image resolution (to be compared to maximum phase shifts of about 1 radian induced by the pressure wave). Moreover, a large asymmetry would have been found between positive and negative pressure swings. None of these two phenomena is observed. Thus transient melting does not occur.
\begin{figure}[ht]
\includegraphics[width=8cm,clip=]{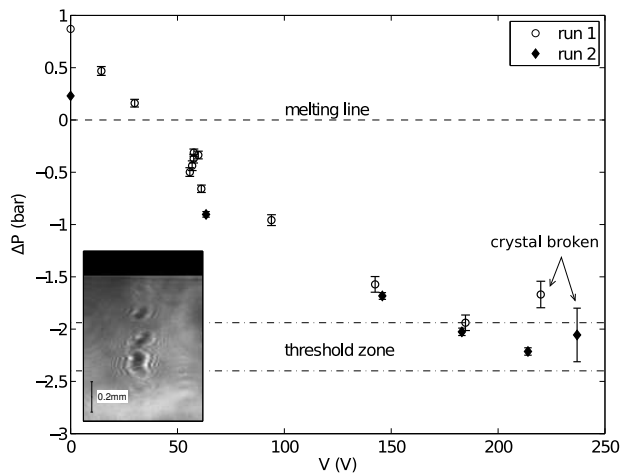}%{merged.eps}
\caption{Minimum depression achieved below the melting pressure $\Delta P$ versus the PZT driving voltage $V$. Diamonds and open circles correspond to different runs with different crystals and different initial pressure offsets $\delta P_0$. Error bars correspond to one standard deviation of the noise on $\delta n$. For the last point on each run, a defect appears at focus (see inset), which is also responsible for the larger noise. The crystal stability limit in $\Delta P$  is between the two dash-dotted lines.} % manip du 4 et 5 mars. Inset: 2011-03-05_09h59m, image_015_0190.dat \deltaP_0=0.87 bar
\label{fig_dnmaxV}
\end{figure}

Is the crystal modified by the pressure wave? Recent experiments\cite{rojas2010,syshchenko2010} show that creation and motion of  dislocations are involved in the response of solid helium-4 even to very low shear strain ($\epsilon \approx10^{-8}$). The temperature used here ($1.1$ K) is significantly higher than the  temperature where the dislocations become free to move. Hence, it is likely that with the strain produced in this experiment (typically 2$\times$10$^{-4}$ at the PZT's surface at $200$~V, 4$\times$10$^{-3}$ at focus) the solid becomes full of dislocations. These dislocations can organize themselves into grain boundaries, leading to a subdivision of the crystal into slightly disoriented crystallites \cite{friedel1964}. We looked for such a crystal modification in two ways. First, we have performed a set of experiments where the driving voltage ($V$) is gradually increased from a reference voltage $V_{\rm ref}=60$~V up to $220$~V. In between each different driving voltage acquisition, a test acquisition at the reference voltage was taken. We then compared the successive test phase maps to the reference map and found them quite similar. For instance, the minimum depressions $\Delta P$ for the reference voltage are within 20\% as shown in Fig.~\ref{fig_dnmaxV} (open circles). For comparison, a misaligned crystal with its $c$-axis at $20^{\circ}$ from the PZT axis or a solid made of a few crystals give very different density maps. As a second test, after sonication at $100$~V, % 19 mars p 105 cahier manip
 the crystal was melted carefully at its coexistence pressure, reversing the procedure used for its growth. The liquid/solid interface reveals the existence of grain boundaries around the focal region. The conclusion is that the local disorientation of the crystal is not sufficient to modify significantly the pressure wave propagation.   

In the same series of measurements, we looked for the maximum achievable depression. It was found that the crystals break when $V$ exceeds a threshold lying between $185$~V and $220$~V. An extended defect with a different density appears at the focal point (see inset of Fig.~\ref{fig_dnmaxV}). The corresponding value $\Delta P$ is between $-1.93$ and $-2.40$~bar. This threshold zone is determined as follows. For a given run at a given temperature, its upper limit is the lowest pressure achieved without breaking the crystal. To estimate its lower limit, we extrapolate {$\Delta P$} from the last two points where the crystal remains stable up to the voltage where the crystal {is broken}. For the two runs of Fig.\ref{fig_dnmaxV}, the given threshold zone is the union of the corresponding threshold zones.

\begin{figure}[ht]
\includegraphics[width=8cm,clip=]{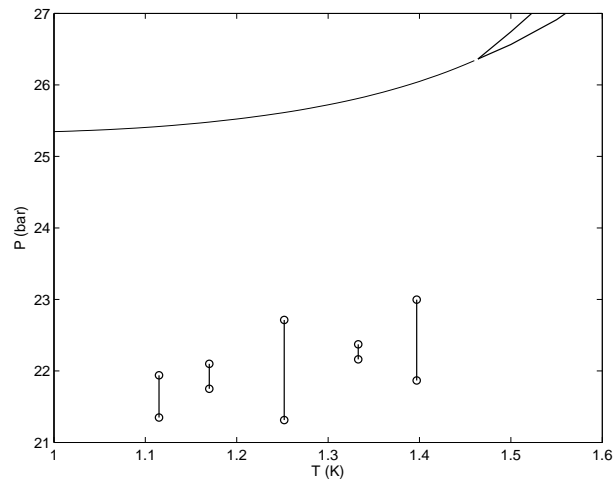}%{merged.eps}
\caption{Threshold zones (crystal stability limit, see text) at different temperatures between 1.1~K and 1.4~K} % manip du 4 et 5 mars. Inset: 2011-03-05_09h59m, image_015_0190.dat \deltaP_0=0.87 bar
\label{fig_fT}
\end{figure}

Once defects are created in the crystal, they disappear in a few seconds after the sound pulse. However, subsequent sound focusing is degraded, indicating a substantial crystal reorganization in the central region. Hence to carry out a new experiment, the crystal has to be melted again and a new one regrown using the procedure shown in Fig.~\ref{chemin}. Clearly the limit encountered here is not the limit of the solid phase discussed in ref.~\cite{maris2009} which is expected at much lower depression ($-35$ bar). In order to get some hints about its origin, 24 different runs were performed varying the temperature, the initial pressure offset $\delta P_0$, the repetition rate of sound pulses. Each run was made with a new crystal. It appears that $\Delta P$ somewhat depends on the crystal, although they are grown in similar conditions. Actually, $\Delta P$ may change by a factor 2 from one sample to another. It was found that $\Delta P$ is sensitive to the emission rate of sound pulses received by the crystal. A smaller rate tends to promote larger $\Delta P$. Surprisingly, we did not find a strong correlation between $\delta P_0$ and $\Delta P$. Finally, $\Delta P$ does not depend much on temperature from $1.1$~K to $1.4$~K as shown in Fig.\ref{fig_fT}. It represents the lowest thresholds achieved for five temperatures in this temperature range.

%It may be related to the bcc phase. Indeed, an extrapolation of the hcp/bcc coexistence line comes close to the observed stability region. The $\mathrm{hcp} \rightarrow \mathrm{bcc}$ transformation is known to exhibit little metastability contrarily to the inverse transformation\cite{grigorev2008}. It could also be a technical limitation of the ultrasound method to create large depressions. Fig.~\ref{fig_dnmaxV} shows that the maximum amplitudes do not scale linearly with $V$ above $60$ V. This reveals some nonlinear process in the production or propagation of the pressure wave, possibly dislocation creation and motion. The temperature dependence of the threshold up to $1.4$ K will probably give a hint for the physics involved.   

From this graph, it seems unlikely that the instability is related to the hcp/bcc coexistence line or its extrapolation in the metastable region. Note that these data were taken at slower sound pulse rates than the data of Fig.\ref{fig_dnmaxV}. This may explain the lower minimum pressures achieved.

A further indication appears in Fig.\ref{fig_dnmaxV}. The values of $\Delta P$ do not scale linearly with $V$ above $V=60$~V. This reveals some non linear process in the production or the propagation of the pressure wave, possibly dislocation creation or motion. The initial dislocation density may most probably vary from one crystal to another; their proliferation might depend on the number of strong sound pulses sent through the sample. If confirmed, this would appear as a technical limitation of the ultrasound technique to achieve larger depressions using a bowl shaped PZT. One can nevertheless try to double $\Delta P$ by using two such PZTs, assembled face to face, to make a nearly complete wave surface.

To summarize, we have demonstrated the production of metastable solid helium between $1.1$~K and $1.4$~K using pressure as a control parameter. For amplitudes up to -2.0~bar below the melting line, the ultrasound technique used does not seem to alter drastically the crystal. An unexpected instability of the solid phase was found. Nevertheless, a metastable region for hcp solid helium-4 is now accessible where crystal properties can  be investigated. Among these, bulk compressibility can be addressed through Brillouin scattering. Vacancy energy will be more difficult to measure, but no less interesting.

%Further investigations are required to determine if this instability is an intrinsic property of the crystal, or an unfortunate consequence of the decompression method used.

%In conclusion, we have demonstrated the production of metastable solid helium using pressure as a control parameter. An unexpected instability of the solid phase was found about $-2.8\pm0.2$ bar below the melting curve which requires further investigations. 
%
%An extension of this work could be the determination of the bulk compressibility in this metastable state by measuring the sound velocity by Brillouin scattering. 

%\begin{table}
%\caption{Table caption.}
%\label{tab.1}
%\begin{center}
%\begin{tabular}{lcr}
%first  & table & row\\
%second & table & row
%\end{tabular}
%\end{center}
%\end{table}

\acknowledgments
We thank S. Balibar and F. Caupin for their help and the technical staff of the Laboratoire Kastler Brossel and of the D\'{e}partement de Physique de l'\'{E}cole Normale Sup\'{e}rieure for a strong support. This research has been supported by ANR grant-05-BLAN-0084-01. 

%\begin{thebibliography}{0}

%\bibitem{b.a}
 % \Name{Author F., Author S. \and Author T.}
  %\REVIEW{Some Rev. A}{69}{1969}{9691}.

%\bibitem{b.b}
  %\Name{Author F. \and Author S.}
  %\Book{Some Book of Interest}
  %\Editor{A. Editor}
  %\Vol{9}
  %\Publ{Publishing house, City}
  %\Year{1939}
  %\Page{666}.

%\bibitem{b.c}
 % \Editor{Editor A.}
  %\Book{Some Book of Interest}
  %\Vol{9}
  %\Publ{Publishing house, City}
  %\Year{1939}
  %\Section{A}.

%\end{thebibliography}
\bibliographystyle{eplbib}
%\bibliography{ref_souris_june2011}

\end{document}